\documentclass[preprint,12pt]{elsarticle}

\usepackage{graphicx}
\usepackage{dcolumn}
\usepackage{bm}
\usepackage{amssymb}
\usepackage{graphicx}

\journal{Commun Nonlinear Sci Numer Simulat}

\begin{document}

\begin{frontmatter}

\title{Mathematical model with autoregressive process for electrocardiogram
signals}

\author{Ronaldo M Evaristo$^{1,2}$, Antonio M Batista$^{2,3,4,5}$, Ricardo L
Viana$^6$, Kelly C Iarosz$^{4,5}$, Jos\'e D Szezech Jr$^{2,3}$, Moacir F de 
Godoy$^7$}
\address{$^1$Instituto Federal de Educa\c{c}\~ao, Ci\^encia e Tecnologia do
Paran\'a, Tel\^emaco Borba, PR, Brazil.}
\address{$^2$P\'os-Gradua\c c\~ao em Ci\^encias, Universidade Estadual de Ponta
Grossa, Ponta Grossa, PR, Brazil.}
\address{$^3$Departamento de Matem\'atica e Estat\'istica, Universidade
Estadual de Ponta Grossa, Ponta Grossa, PR, Brazil.}
\address{$^4$Instituto de F\'isica, Universidade de S\~ao Paulo, S\~ao Paulo, 
SP, Brazil.}
\address{$^5$Institute for Complex Systems and Mathematical Biology, Aberdeen,
Scotland, UK.}
\address{$^6$Departamento de F\'isica, Universidade Federal do Paran\'a ,
Curitiba, PR, Brazil.}
\address{$^7$Faculdade de Medicina de S\~ao Jos\'e do Rio Preto, S\~ao Jos\'e
do Rio Preto, SP, Brazil.}

\cortext[cor]{Corresponding author: antoniomarcosbatista@gmail.com}

\date{\today}

\begin{abstract}
The cardiovascular system is composed of the heart, blood and blood vessels. 
Regarding the heart, cardiac conditions are determined by the electrocardiogram,
that is a noninvasive medical procedure. In this work, we propose
autoregressive process in a mathematical model based on coupled differential
equations in order to obtain the tachograms and the electrocardiogram signals
of young adults with normal heartbeats. Our results are compared with
experimental tachogram by means of Poincar\'e plot and dentrended fluctuation
analysis. We verify that the results from the model with autoregressive process
show good agreement with experimental measures from tachogram generated by
electrical activity of the heartbeat. With the tachogram we build the
electrocardiogram by means of coupled differential equations.
\end{abstract}

\begin{keyword}
heartbeat \sep autoregressive model \sep electrocardiogram
\end{keyword}

\end{frontmatter}


\section{Introduction}

The cardiovascular system (CVS) is responsible for supplying the human organs
with blood. It is composed by the heart, the arteries, and the veins. The heart
has as function to pump blood throughout the body, that is realised by means of
contractions \cite{mohrman14}. The human heart beats an average $72$ beats per 
minute and pumps 0.07 liters of blood per beat \cite{curtis89,hall11}. The
contraction and relaxation of the heart is obtained by a single cycle of the
electrocardiogram signal (ECG), namely the ECG records of the electrical
activity of the heart. Waller in 1887 \cite{sykes87} measured for the first
time the electrical activity from the heart, and the first practical
electrocardiograph was invented by Einthoven in 1901 \cite{ruiz08} that it was
used as a tool for the diagnosis of cardiac abnormalities.

In the recent past, several theoretical investigations pertaining to CVS have 
been carried out to analyse electrocardiogram signal 
\cite{gaetano09,kudinov15,schenone16}. Mathematical models have been developed 
to understand physiological function and disfunction in CVS. A mathematical 
model which have been used to generate ECG signals is the Van der Pol oscillator
\cite{pol26}. Gois and Savi considered three modified Van der Pol oscillators 
connected by time delay coupling to describe heart rhythm behaviour 
\cite{gois09}. The coupled Van der Pol oscillators was also used in studies 
about the control of irregular behaviour in pathological heart rhythms 
\cite{ferreira14}. McSharry and collaborators \cite{mcsharry03} introduced a 
dynamical model to describe generating synthetic electrocardiogram signals.
This model is based on a set of three ordinary differential equations in that
it is incorporated the respiratory sinus arrhythmia (RSA) by means of a bimodal
power spectrum consisting of the sum of two Gaussian distributions.

In this work, instead two Gaussian distributions we propose an autoregressive
(AR) process for the RSA to obtain the tachograms and, consequently, the ECGs
of young adults with normal heartbeats. AR model can be used to quantify gains
and delays by which cardiac interval, lung volume, blood pressure and
sympathetic activity affect each other \cite{cohen02}. Cardiologists utilise AR
model when they are interested in fit tachograms through mathematical
regressions softwares. The tachogram is a signal that allows the study of heart
rate variability (HRV) \cite{yasuma04,acharya06}. Boardman and collaborators
\cite{boardman02} study autoregressive model for the HRV. They found the
optimum order of autoregressive model which can be used for spectral analysis
of short segments of tachograms. We compare the results obtained from two
Gaussian distributions and AR process with experimental tachograms of healthy
young adults. To do that, we use as diagnostic tools the Poincar\'e plot
\cite{guzik07} and detrended fluctuation analysis (DFA) \cite{peng94}. We have
verified that the result with AR process agrees with the experimental tachogram
more closely than the result with two Gaussian distributions. This way, we
generate the ECG by means of coupled differential equations considering the AR
process to obtain the tachogram and the frequency of the ECG. The frequency is
an important parameter in the mathematical model for the ECG signal, and
variation in the frequency produces variation in the times elapsed between
successive heartbeats.

This article is organised as follows: in Section 2 we introduce the ECG model
with autoregressive process for tachograms, in Section 3 we compare our results
with experimental data, and in the last Section we draw our conclusions.


\section{The mathematical model}

\begin{figure}[htbp]
\begin{center}
\includegraphics[height=5.5cm,width=11cm]{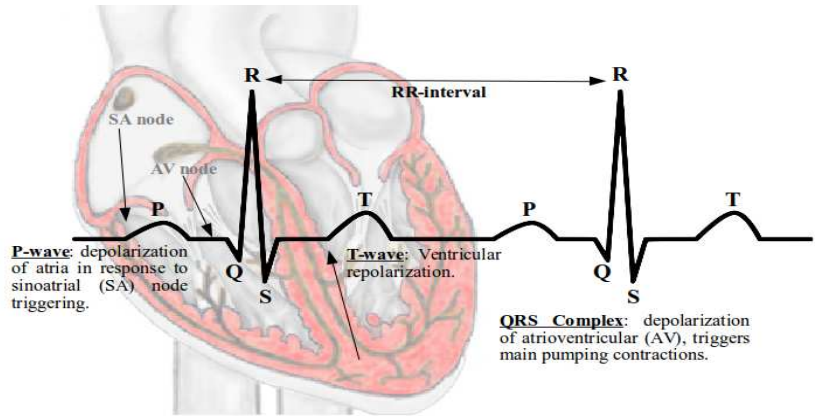}
\caption{(Colour online) ECG following a normal hearbeat.}
\label{fig1}
\end{center}
\end{figure}

The ECG is a noninvasive method used to measure electrical activity of the 
heart through electrodes placed on the surface of the skin. Figure \ref{fig1}
shows the relationship between the cardiac conduction and the ECG. In the
sinoatrial node (SA), known as natural pacemaker, the heartbeat starts. The
atrioventricular node (AV) is responsible for the passage of electrical
signals from the atria to the ventricles. At last, the signal arrives at the
Purkinje fibers and makes the heart contract to pump blood, where the R-peak
occurs. The time between successive R-peaks is the RR-interval and the series
of RR-intervals is known as RR tachogram.

McSharry and collaborators \cite{mcsharry03} argued that the heartbeat can
be described by means of three coupled ordinary differential equations with the
inclusion of RSA  at the high frequencies ($f_{\rm RSA}$) and Mayer waves (MW) at
the low frequencies ($f_{\rm MW}$). The equations are given by
\begin{eqnarray}\label{eqgaus}
{\dot x} & = & \alpha x-\omega y,\nonumber \\
{\dot y} & = & \alpha y+\omega x, \\
{\dot z} & = & z_0-z-\sum_i a_i\Delta\theta_i
               {\rm e}^{-\frac{\Delta\theta_i^2}{2b_i^2}}, \nonumber
\end{eqnarray}
where $i\in\{P,Q,R,S,T\}$, $\alpha=1-\sqrt{x^2+y^2}$,
$\Delta\theta_i=\theta-\theta_i$ (mod $2\pi$), $\theta={\rm atan}2(y,x)$
($-\pi\leq {\rm atan}2(y,x)\leq \pi$), $z_0(t)=A\sin(2\pi f_{\rm RSA}t)$, and
$A=0.15$mV. Visual analysis of a ECG from a normal individual is used to obtain
the times, as well as the angles $\theta_i$, the $a_i$ and $b_i$ values for the
PQRST points. The parameters values are given in Table \ref{pareq} according to
Reference \cite{mcsharry03}.

\begin{table}
\centering
\caption{Parameters for Equation (\ref{eqgaus}).}
\label{pareq}
\begin{tabular}{ccccc}
\hline
index (i) & time (s) & $\theta_i$ (rad) & $a_i$ & $b_i$ \\ \hline
P  &  -0.2  & $-\pi/3$  &  1.2  &  0.25 \\
Q  &  -0.05 & $-\pi/12$ &  -5.0 &  0.1  \\
R  &   0    &   0               &  30.0 &  0.1  \\
S  &   0.05 &  $\pi/12$ & -7.5  &  0.1  \\
T  &   0.3  &  $\pi/2$  &  0.75 &  0.4  \\
\hline
\end{tabular}
\end{table}

\begin{figure}[htbp]
\begin{center}
\includegraphics[height=7cm,width=8cm]{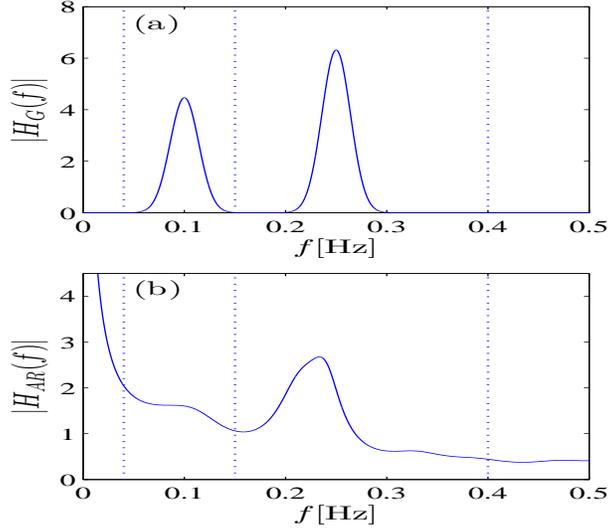}
\caption{Power spectrum from (a) Equation (\ref{eqbimodal}) and (b)
Equation (\ref{HAR}).}
\label{fig2}
\end{center}
\end{figure}

The frequency $\omega(t)$ controls the variations in the RR-intervals, and it
is given by
\begin{equation}\label{omega}
\omega(t)=\frac{2\pi}{r(t)},
\end{equation}
where $r(t)$ is the continuous version of the $r(n)$ time series which is
obtained from the inverse discrete-time Fourier transform (DTFT)
\cite{oppenheim10,proakis07} of the power spectrum
\begin{equation}\label{eqbimodal}
|H_G(f)|^2=\frac{\sigma^2_{\rm MW}}{\sqrt{2\pi c_{\rm MW}^2}}
           \exp \frac{(f-f_{\rm MW})^2}{2c^2_{\rm MW}}+
           \frac{\sigma^2_{\rm RSA}}{\sqrt{2\pi c_{\rm RSA}^2}}
           \exp \frac{(f-f_{\rm RSA})^2}{2c^2_{\rm RSA}},
\end{equation}
with phase randomly distributed from $0$ to $2\pi$ \cite{mcsharry03}. The
tachogram exhibits similarity with a real one when the phase is randomly
distributed. Figure \ref{fig2}(a) exhibits the power spectrum $|H_G(f)|$ for
$f_{\rm MW}=0.1$Hz, $f_{\rm RSA}=0.25$Hz, $c_{\rm MW}=0.01$, $c_{\rm RSA}=0.01$, and
$\sigma^2_{\rm MW}/\sigma^2_{\rm RSA}=0.5$. The values of the frequencies $f_{MW}$
and $f_{RSA}$ are motivated by the power spectrum of a real RR tachogram. The
$f_{\rm MW}$ value is approximately equal to $0.1$ in humans, and it is related
to arterial pressure occurring spontaneously in conscious subjects. RSA is
characterised by the presence of oscillations in the tachogram considering the
parasympathetic activity. It is synchronous with the breathing rate which for
normal subjects is equal to $15$ breaths per minute, i.e., $f_{\rm RSA}=0.25$Hz.
The spectrum has a bimodal form, where one peak is located in the low frequency
range $0.04\leq f< 0.15$Hz and the other is located in the high frequency range
$0.15\leq f\leq 0.4$Hz. These two bands appear due to the effects of both Mayer
waves and RSA. The tachogram $r(n)$ is generated by the inverse DTFT from the
power spectrum $|H_G(f)|$ with random phase. To obtain the continuous signal
$r(t)$, first we increase the sample rate of $r(n)$ to the same sample rate of
the desired ECG by interpolation \cite{mcsharry03,oppenheim10,proakis07}. Then,
a R-peak detection algorithm \cite{mcsharry03} is applied to the interpolated
signal to determine $r(t)$, as a result $\omega(t)$ (Eq. \ref{omega}) is
calculated and the ECG is built by means of Equation (\ref{eqgaus}).

In this work, in order to model electrocardiogram signals we propose an
autoregressive (AR) stochastic process \cite{boardman02} to determine
$\omega(t)$. The AR process of order $p$ is defined as
\begin{equation}
R(n)=\sum_{l=1}^p d_lR(n-l)+\epsilon(n),
\end{equation}
where $\epsilon(n)$ is a white noise with zero mean and unit variance. Boardman
and collaborators \cite{boardman02} found that $p=16$ is an optimum order of
autoregressive model for heart rate variability. The AR power spectrum density
is
\begin{equation}\label{HAR}
|H_{AR}(f)|=\frac{1}{\left|1-\sum_{l=1}^p d_l{\rm e}^{-{\rm i}2\pi fl}\right|},
\end{equation}
with the coefficients values given in Table \ref{tabd}. The coefficients values
for the AR power spectrum density have been adjusted to be used for the set of
data that we obtained from healthy individuals. The presence of arrhythmia can
influence the coefficients values, and as a consequence it would be necessary
to calculate the new coefficients values.

\begin{table}
\centering
\caption{Coefficients values for the AR power spectrum density (Eq. \ref{HAR}).}
\label{tabd}
\begin{tabular}{llll}
\hline
$d_{1}=-0.9099$ & $d_{2}=0.5188$   & $d_{3}=-0.2840$ & $d_{4}=-0.2063$ \\ 
$d_{5}=0.0382$  & $d_{6}=0.0709$   & $d_{7}=0.0305$  & $d_{8}=-0.1533$ \\ 
$d_{9}=0.0009$  & $d_{10}=-0.0070$ & $d_{11}=-0.0218$ & $d_{12}=0.0043$ \\ 
$d_{13}=0.0316$ & $d_{14}=0.0155$  & $d_{15}=-0.0591$ & $d_{16}=0.0252$ \\ \hline
\end{tabular}
\end{table}

Figure \ref{fig2}(b) shows the power spectrum calculated from the tachogram
generated by means of the AR process. In placed of the two separated Gaussian
distributions, the AR process produces a damped in the power spectrum, and as
consequence the separation between the frequency components cannot be exactly
identified, as in real situation. This way, with the tachogram $r(n)$ we find
$r(t)$ (interpolation followed by R-peak detection) and $\omega(t)$ to build
the ECG using Equation (\ref{eqgaus}).

\section{Results and discussions}

\begin{figure}[htbp]
\begin{center}
\includegraphics[height=6.5cm,width=8cm]{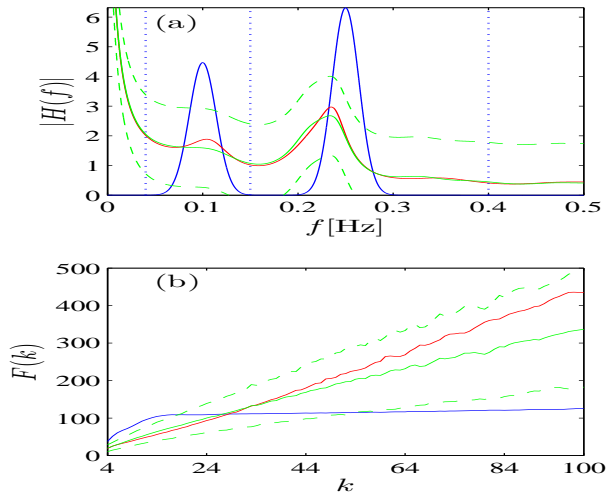}
\caption{(Colour online) (a) Power spectrum and (b) dentrended fluctuation
analysis (DFA) for Gaussian distributions (blue), AR process (red), and
experimental data (green). The green dashed lines exhibit the confidence
interval of the experimental power spectrum that are calculated from $62$
experimental tachograms.}
\label{fig3}
\end{center}
\end{figure}

In this work, the power spectrum is considered to obtain the theoretical
tachogram and consequently the frequency (Eq. \ref{omega}) that is used 
in Equation (\ref{eqgaus}) to build the ECG. We generate $124$ theoretical ECG
signals, being $62$ from the Gaussian spectrum and $62$ from the AR spectrum.
Then, we obtain their respective tachograms and compare them with $62$
experimental tachograms collected from healthy adults. The experimental
protocol consisted of $20$min of monitoring of the heartbeat frequency in
resting from patients without sound and visual stimulations in a supine rest
position. The heartbeats were recording with a sample rate of $1000$Hz, and
$1000$ RR intervals were analysed. The experimental tachograms were filtered to
remove ectopic beats and noise effects. We calculate the power spectrum from
Gaussian distributions (blue), AR process (red), and experimental data (green),
as shown in Figure \ref{fig3}(a). The green dashed lines exhibit the confidence
interval of the experimental power spectrum that are calculated from $62$
experimental tachograms.

With regard to Figure \ref{fig3}(a), we see that the power spectrum from AR
process has a shape closer the experimental result than the power spectrum from
the Gaussian distributions. This way, in order to verify the agreement among
the experimental tachogram and the tachograms obtained from theoretical power
spectra we have utilised the dentrended fluctuation analysis (DFA)
\cite{hu01,chen02}. DFA yields a fluctuation function $F(k)$ as a function of
$k$, given by \cite{peng94}
\begin{equation}
  F(k)=\sqrt{\frac{1}{N}\sum_{n=0}^{N-1}[r_I(n)-r_k(n)]^2},
\end{equation}
where $N=1000$, $r_I(n)=\sum_{l=0}^{n}r(l)$ is the cumulative sum of the $r(n)$,
$k$ is the box size that partitions the time interval of the tachogram, and
$r_k(n)$ is the local trend in each box. DFA is a nonlinear dynamical analysis
that have been used for the understanding of biological systems \cite{peng94}.
Moreover, the DFA allows the detection of long-range correlations embedded in a
patchy landscape.

Figure \ref{fig3}(b) shows the mean DFAs for $4<k<100$, where linear regressions
present slopes equal to $0.1955\pm 0.0150$, $0.7034\pm 0.0686$, and
$0.6817\pm 0.2448$ for Gaussian distributions, AR process and experimental
data, respectively. The experimental data are collected from $62$ healthy young
adults. Each one with length equal to $1000$ heartbeats without trend removal.
In general, ectopic beats or noise effects are excluded of time series through
filters \cite{santos13}. As a result, the DFA for the tachogram generated by
the experimental data and AR process are in close agreement with each other.
Whereas DFA for the Gaussian distributions exhibits a good agreement only for
$k<20$.

\begin{figure}[htbp]
\begin{center}
\includegraphics[height=7.5cm,width=12cm]{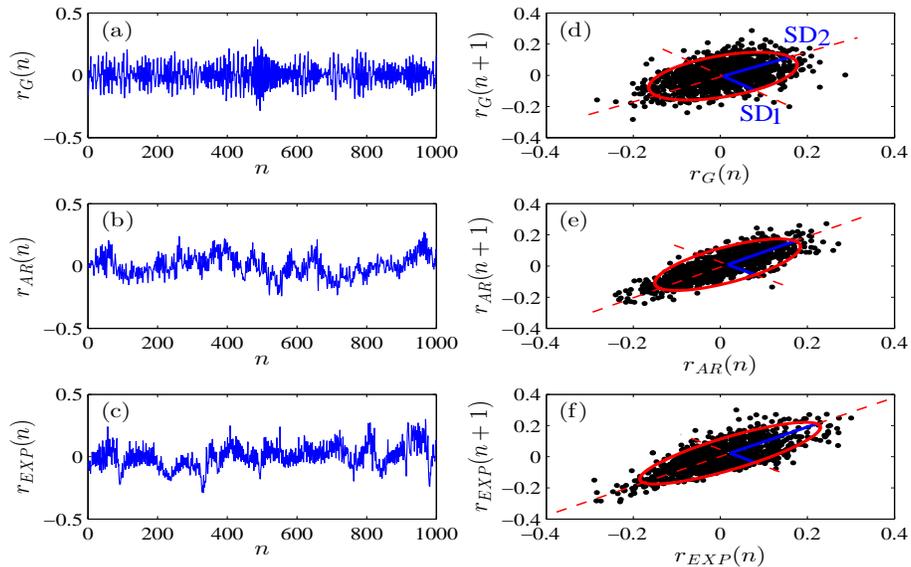}
\caption{(Colour online) Tachograms generated by (a) the sum of two Gaussian
distributions, (b) the AR process, (c) experimental data, and respective
Poincar\'e plots in (d), (e), and (f).}
\label{fig4}
\end{center}
\end{figure}

Through the Gaussian power spectrum with phases randomly distributed between
$0$ and $2\pi$ we build a ECG and extract the tachogram $r_G(n)$ shown in
Figure \ref{fig4}(a). A tachogram extracted of ECG generated by the AR
spectrum $r_{AR}(n)$ is illustrated in Figure \ref{fig4}(b) and an experimental
tachogram $r_{EXP}(n)$ of the healthy adult is in Figure \ref{fig4}(c). In
Figures \ref{fig4}(d), \ref{fig4}(e), and \ref{fig4}(f) we calculate the
respective Poincar\'e plots. The Poincar\'e plot is a visualising technique to
analyse RR intervals, where it is computed the standard deviation of points
perpendicular to the axis (SD1) and points along (SD2) the axis of
line-of-identity. Table \ref{tabSD} exhibits the mean SD1 and SD2 values of the
tachograms shown in Figure \ref{fig4}. Comparing the Poincar\'e plots we see 
that both SD1 and SD2 for the AR process agree with the experimental results 
better than the method based on the Gaussian distribution. 

In addition, we calculate the $p$-values according to the two-sided Wilcoxon
rank sum test \cite{wilcoxon45} of the SD1 and SD2 time series from the
simulated and experimental data. This statistical test verifies if two paired
time series have the same distribution when the data cannot be assumed to be
normally distributed. We find that the $p$-values of SD1 and SD2 are greater
than $0.05$ for the AR process, consequently the time series of SD1 and SD2 in
the AR process and the experimental data have the same distributions. However,
the same does not happen for the Gaussian distributions, where the $p$-values
are less than $0.05$.

\begin{table}
\centering
\caption{Mean SD1 and SD2 values for the theoretical and experimental tachograms
with $p$-values of the Wilcoxon rank sum test.}
\label{tabSD}
\begin{tabular}{ccccc}
\hline
RR-intervals & $r_G(62)$ [A] & $r_{AR}(62)$ [B] & $r_{EXP}(62)$ [C] & $p$-value
\\ \hline
SD1 [ms] & $63.9\pm 1.9$  & $32.8\pm 10.5$ & $38.8\pm 19.6$ &
$0.0$ [A$\times$C] \\
         &                &                &                &
$0.2$ [B$\times$C] \\
\hline
SD2 [ms] & $103.4\pm 2.6$ & $82.0\pm 26.2$ & $79.0\pm 28.5$ &
$0.0$ [A$\times$C]\\ 
         &                &                &                &
$0.6$ [B$\times$C]\\  
\hline
\end{tabular}
\end{table}

\begin{figure}[htbp]
\begin{center}
\includegraphics[height=5cm,width=8cm]{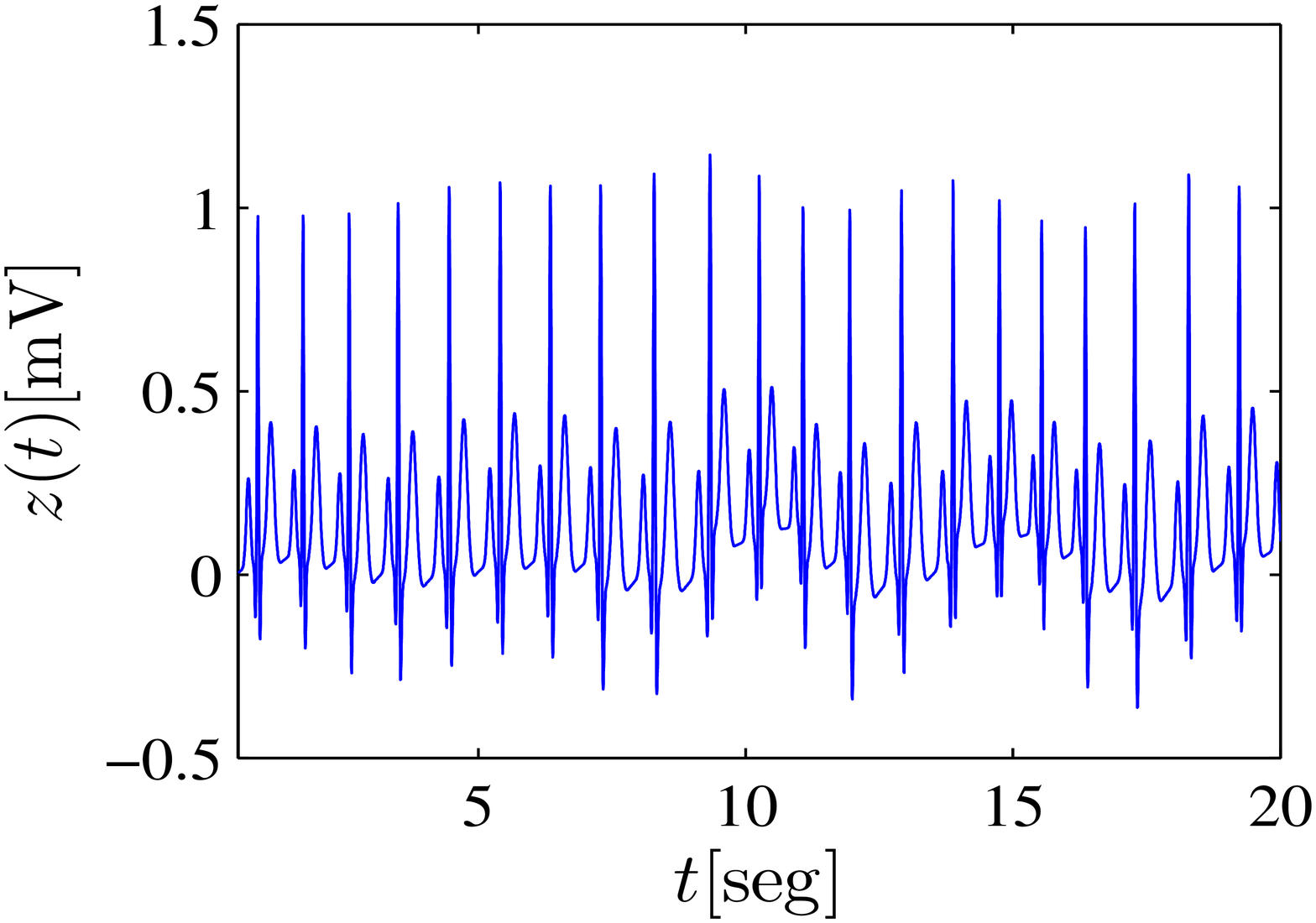}
\caption{ECG signal generated by means of Equation (\ref{eqgaus}) considering
AR process.}
\label{fig5}
\end{center}
\end{figure}

All in all, we obtain the tachogram $r(n)$ (discrete-time) by means of the AR
process. Then, an interpolation followed by a R-peak detection allows us to
determine the signal $r(t)$ (continuous-time). As a result, we calculate
$\omega(t)$ and solve the ordinary differential equations to build the ECG 
signal. We use the fourth order Runge-Kutta method to solve the ordinary
differential equations, where we consider a fixed time step $\Delta t=1/f_s$
and sampling frequency $f_s=256$Hz. Figure \ref{fig5} shows the ECG signal in
the time interval $0\leq t\leq 20$s, where $z(t)$ yields a synthetic ECG with a
realistic PQRST morphology according to Figure \ref{fig1}.


\section{Conclusions}

In conclusion, we have studied a mathematical model given by coupled
differential equations that describes electrocardiogram signals and it was
proposed by McSharry and collaborators \cite{mcsharry03}. In the original
model, the frequency is calculated from a power spectrum with two Gaussian
distributions that incorporates both respiratory sinus arrhythmia and Mayer
waves. The use of two Gaussian distributions is in disagreement with our
experimental data obtained from healthy adults, due to the fact that the two
distributions are not well separated in the power spectrum. This way, we
propose to calculate the frequency by means of the AR process. We believe that
our model also allows the study of respiratory sinus arrhythmia.

We verify that the power spectrum from AR process has a good agreement with
the power spectrum from experimental data. We have also compared the tachograms
generated from Gaussian distributions and AR process with the experimental
tachogram using DFA and Poincar\'e plot. As a result, in both DFA and Poincar\'e
plot, the tachogram generated considering the AR process is in closer agreement
with experimental data than the two Gaussian. As a consequence, with the
tachogram, the frequency is calculated and the ECG signal can be built
utilising coupled differential equations with AR process.

We believe that the mathematical model with autoregressive process constitutes
an important step toward developing strategies to simulate electrocardiogram
signals. We have considered many experimental tachograms from healthy adults to
obtain the parameters, this way our simulations allow to obtain and also to do
forecast of electrocardiogram signals of individuals in the same situation
analysed in this work. We have tested our results using experimental tachograms
obtained from $62$ healthy patients without sound and visual stimulations in a
supine rest position. In future works, we plan to do the same analyse
considering patients with different clinical conditions.


\section*{Acknowledgements}

This study was possible by partial financial support from the following
agencies: Funda\-\c c\~ao Arauc\'aria, Brazilian National Council for
Scientific and Technological Development (CNPq), Coordination for the
Improvement of Higher Education Personnel (CAPES), and S\~ao Paulo Research
Foundation (FAPESP) process numbers 2011/19296-1, 2015/07311-7, and
2016/16148-5.


\end{document}